# Complex Permittivity Measurements at Variable Temperatures of Low Loss Dielectric Substrates Employing Split Post and Single Post Dielectric Resonators

Janina E. Mazierska[ab], Jerzy Krupka[c], Mohan V. Jacob[a] and Dimitri O. Ledenyov[a]

[a]James Cook University, Electrical and Computer Engineering, Townsville, Qld 4811, Australia
[b]Massey University, Institute of Information Sciences and Technology, PB 11222, Palmerston North, New Zealand
[c]Instytut Mikroelektroniki i Optoelektroniki Politechniki Warszawskiej, Koszykowa 75, 00-662 Warszawa, Poland

*Abstract* – **A split post dielectric resonator in a copper enclosure and a single post dielectric resonator in a cavity with superconducting end-plates have been constructed and used for the complex permittivity measurements of single crystal substrates. (La,Sr)(Al,Ta)$O_3$, $LaAlO_3$, MgO and quartz substrates have been measured at temperatures from 20 K to 300 K in the split post resonator and from 15 K to 80 K in the single post resonator. The $TE_{01\delta}$ mode resonant frequencies and unloaded $Q_o$-factors of the empty resonators at temperature of 20K were: 9.952GHz and 25,000 for the split post resonator and 10.808GHz and 240,000 for the single post resonator respectively.**

## I. INTRODUCTION

Precise microwave characterization of low loss dielectric substrates represents still an interesting and challenging task. For cylindrical samples the whispering gallery mode (WGM) technique allows for measurements of arbitrary low loss (by choosing a mode for which all losses in a resonator are negligible apart from the dielectric loss). Unfortunately there is no equivalent of the WGM technique that can be applied to planar dielectrics. Hence low loss dielectric substrates can be characterized at best using the split post dielectric and the split cavity resonators. Less precise methods include the microstrip and stripline resonator techniques and using data measured for cylindrical samples of the same material.

The split-post dielectric resonator (SPDR) technique has proved to be useful for measurements of the complex permittivity of dielectric laminar specimens at frequency range 1-30 GHz [1-6]. Typically SPDR measurement fixtures are designed for measurements at room temperature or in a narrow range of temperatures close to 300K. We have constructed a SPDR to operate in a wide range of temperatures; from 20K to 400K. The constructed resonator was used for measurements of low loss single crystal planar dielectrics and verified with a single post dielectric resonator of higher resolution.

## II. SPLIT POST AND SUPERCONDUCTING SINGLE POST DIELECTRIC RESONATORS

The fabricated SPDR is schematically shown in Fig. 1. It employs BMT dielectric resonators on quartz support and contains no plastic parts.

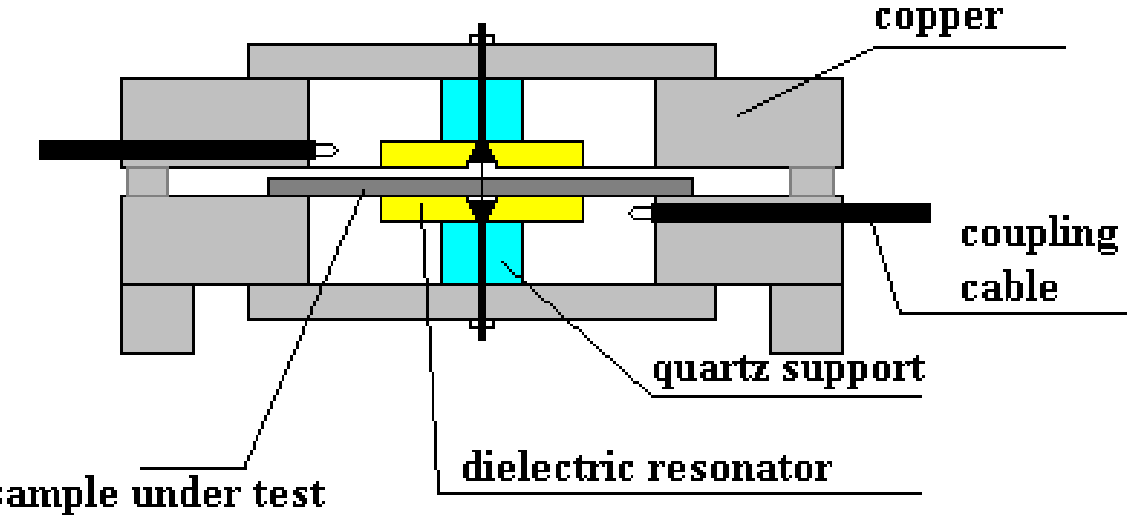

Fig. 1. Schematic diagram of the Split Post resonator

The empty resonator exhibited the resonant frequency of the $TE_{01\delta}$ mode of 9.952 GHz at 20 K and the $Q_o$-factor of 25,000. The measured temperature dependences of the $Q_o$-factor and $f_{res}$ of this fixture from 25K to 300K are presented in Fig. 2. The electromagnetic analysis of the designed SPDR has been based on the rigorous full wave Rayleigh Ritz technique [5, 7]. The real part of permittivity of a substrate under test is calculated as:

$$\varepsilon_r' = 1 + \frac{f_o - f_{res}}{h f_o K_\varepsilon(\varepsilon_r', h)} \qquad (1)$$

where $f_o$ and $f_{res}$ are resonant frequencies of the empty resonator and with a sample respectively, h is the substrate's thickness and $K_\varepsilon$ is a function of $\varepsilon_r$' and *h*. The constant $K_\varepsilon$ is pre-computed and tabulated for a number of $\varepsilon_r$' and h. Interpolation is then used to compute $K_\varepsilon$ for specific permittivity and thickness values.

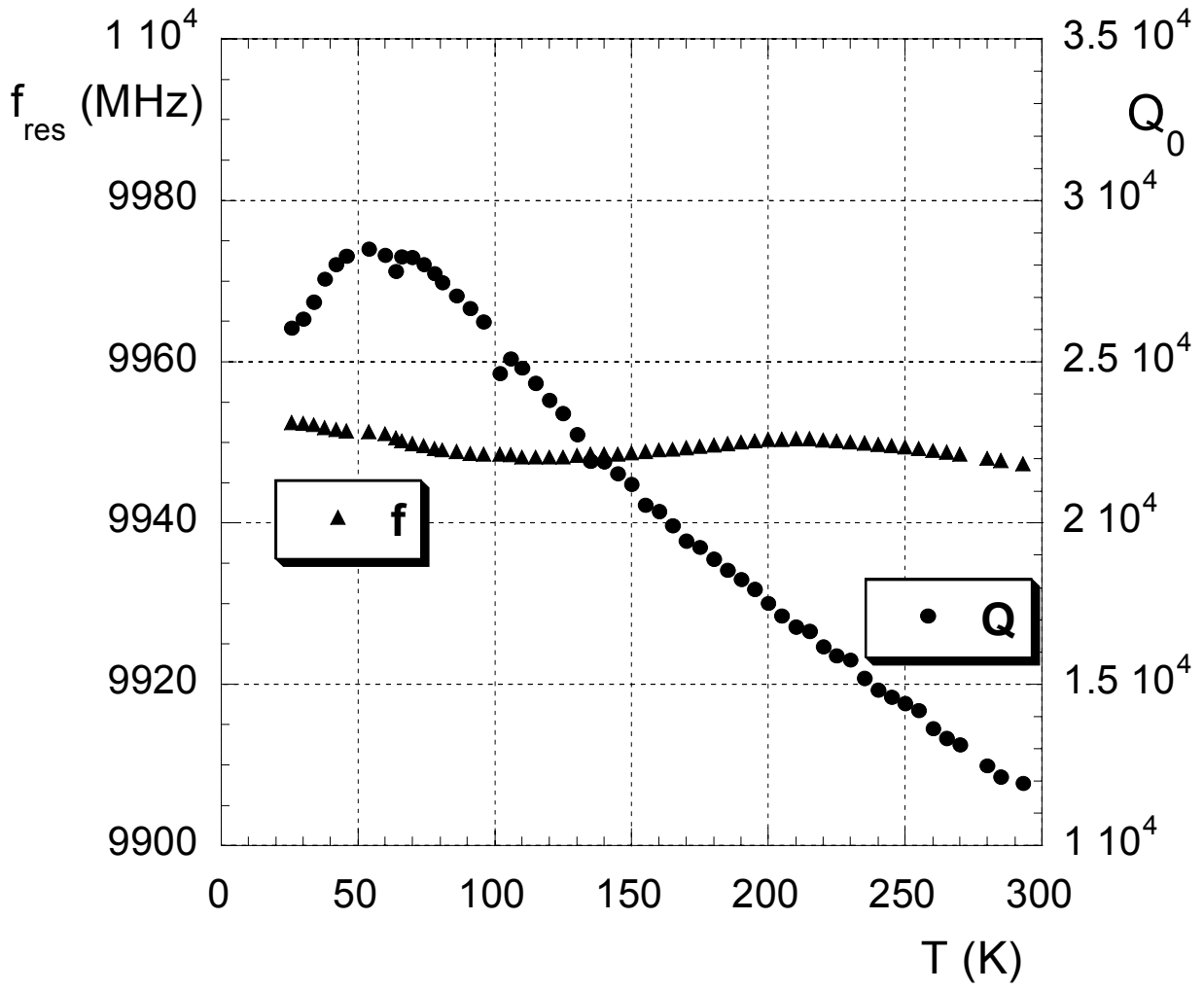

Fig.2. Unloaded $Q_o$ and $f_{res}$ of the Split Post resonator (SPDR)

The loss tangent of a tested substrate is determined from the measured unloaded $Q_o$-factor of the post resonator with the tested substrate using the loss equation as in [7]:

$$\tan\delta = (Q_o^{-1} - Q_{DR}^{-1} - Q_c^{-1}) / p_{es} \qquad (2)$$

where $Q_{DR}$ represents dielectric losses of the resonator:

$$Q_{DR} = Q_{DR0}(f_0/f_s)(p_{eDR0}/p_{eDR}) \qquad (3)$$

where $Q_{DR0}$ is the dielectric loss in the empty fixture, $p_{eDR}$ and $p_{eDR0}$ are electric energy filling factors with the sample and empty respectively,

$Q_c$ represents the conductor losses:

$$Q_c = Q_{c0} K_1(\varepsilon_r', h) \qquad (4)$$

where $Q_{c0}$ describes conductor losses of the empty resonator, and $K_1$ is a function of $\varepsilon_r$' and *h*,

$p_{es}$ is the electric energy filling factor of the sample:

$$p_{es} = h\varepsilon_r' K_p(\varepsilon_r', h) \qquad (5)$$

where $K_p$ is a function of $\varepsilon_r$' and *h*.

Again, the functions $K_1$, and $K_p$, are pre-computed and tabulated for a number of $\varepsilon_r$' and *h* and interpolation is used to compute their values for specific $\varepsilon_r$' and *h*.

The designed SPDR allows measurements of the perpendicular component of the real permittivity with uncertainty smaller than 0.5% [5], and loss tangent with resolution of approximately $2\times10^{-5}$ at 20K. Such resolution may be not sufficient for measurements of some single crystal materials, eg used as substrates for deposition of High Temperature Superconducting thin films (HTS) at cryogenic temperatures. To verify measured tanδ of such very low loss dielectrics, the same substrates are measured in a single post dielectric resonator with superconducting end-plates (Su PDR) [8] schematically shown in Fig. 3. The Su PDR operates at temperatures up to 80K and exhibits the resonant frequency of 10.808GHz and the $Q_o$ factor of 240,000 at 20K as presented in Fig. 4. Due to obtained $Q_o$ values the resolution in tanδ measurements is very high; reaching $2\times10^{-6}$ at the lowest temperature. The Su PDR allows for a significant increase in the loss tangent resolution of measurements at cryogenic temperatures.

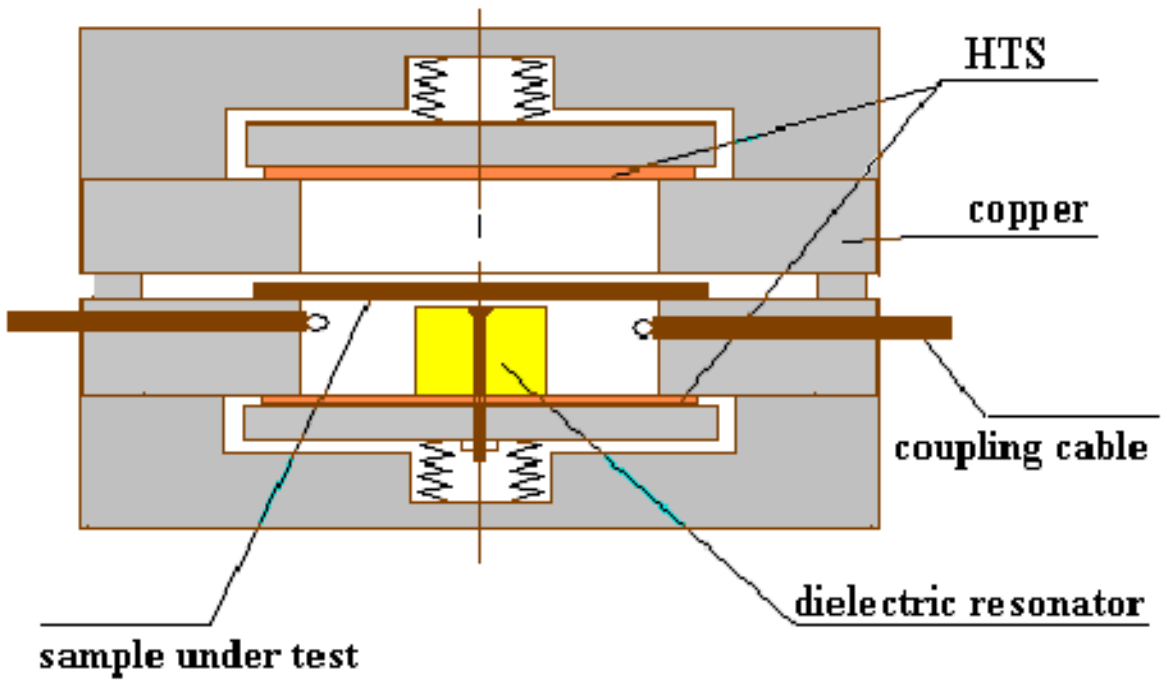

Fig.3. Schematic diagram of Single Post dielectric resonator with superconducting end-plates (Su PDR)

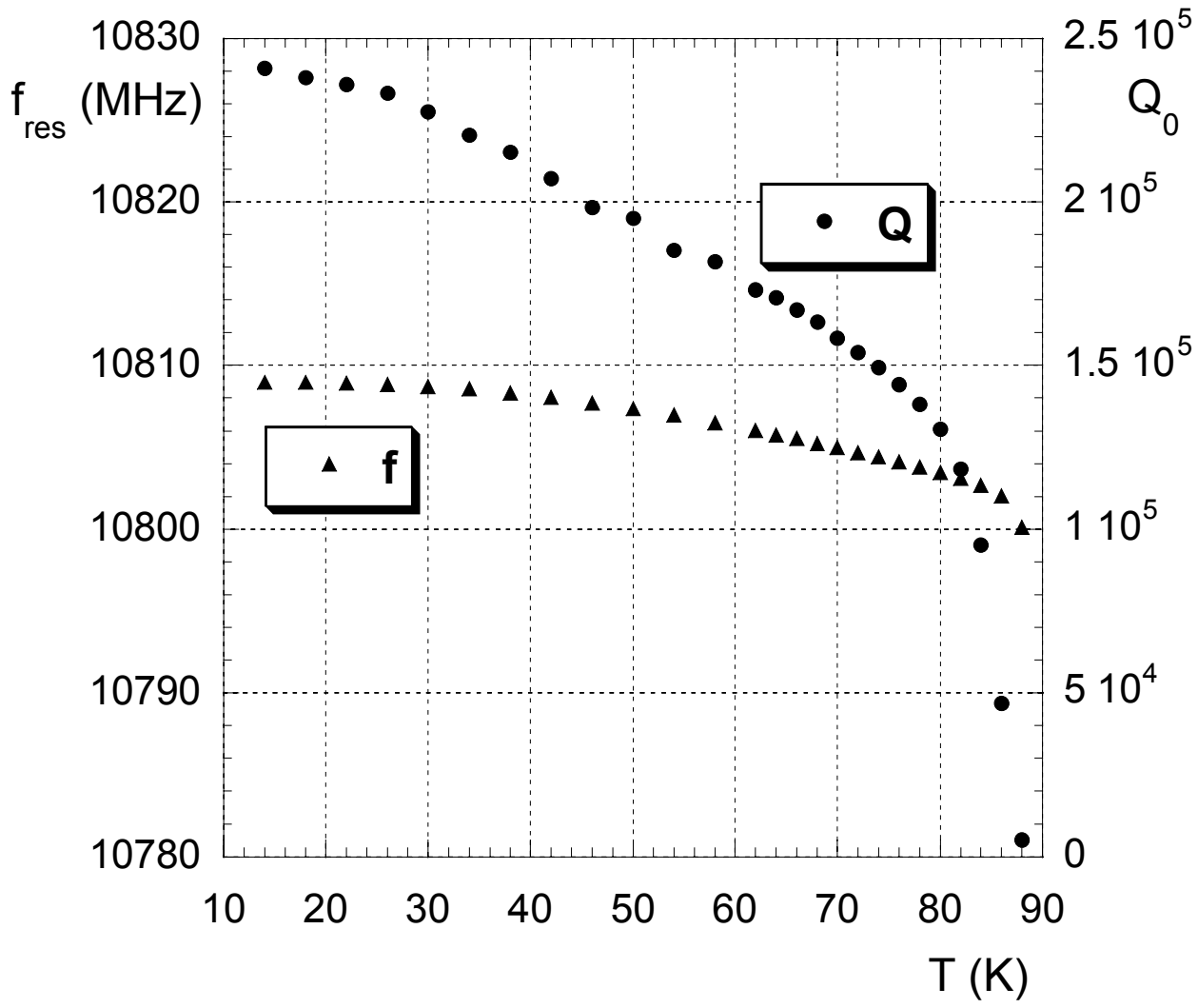

Fig 4. $Q_o$-factor and $f_{res}$ of Su PDR resonator

The complex permittivity of a substrate under test is calculated for the Su PRD in a similar way as for the SPDR. The combined pair of constructed resonators enables relatively precise microwave characterisation of substrates for HTS films and other low loss substrates in a wide range of temperatures from 20K to 400K.

## III. EXPERIMENTS

The constructed Split Post and the Superconducting Single Post Dielectric resonators were used to measure the perpendicular component of the complex permittivity of (La,Sr)(Al,Ta)$O_3$, (LSAT), $LaAlO_3$ (LAO), MgO and quartz substrates. Thickness of the samples was 500μm, 513μm, 508μm and 400μm respectively. The measurement system consisted of a Vector Network Analyser (HP 8722C), Temperature Controller (Conductus LTC-10), Vacuum Dewar, Close Cycle cryocooler (APD-HC4) and a PC. Measurements were based on the simplified Transmission Mode Q Factor data processing technique [9, 10] of $S_{21}$, $S_{11}$ and $S_{11}$ parameters to compute resonant frequencies and unloaded $Q_o$-factors of the empty resonators and the resonators with substrates under test, for each temperature. The $\varepsilon_r$' and tanδ of the samples were evaluated based of the Rayleigh-Ritz analysis as described in Section II. Results of $\varepsilon_r$' dependence for LSAT, LAO and MgO substrates are shown in Fig. 5-7 respectively.

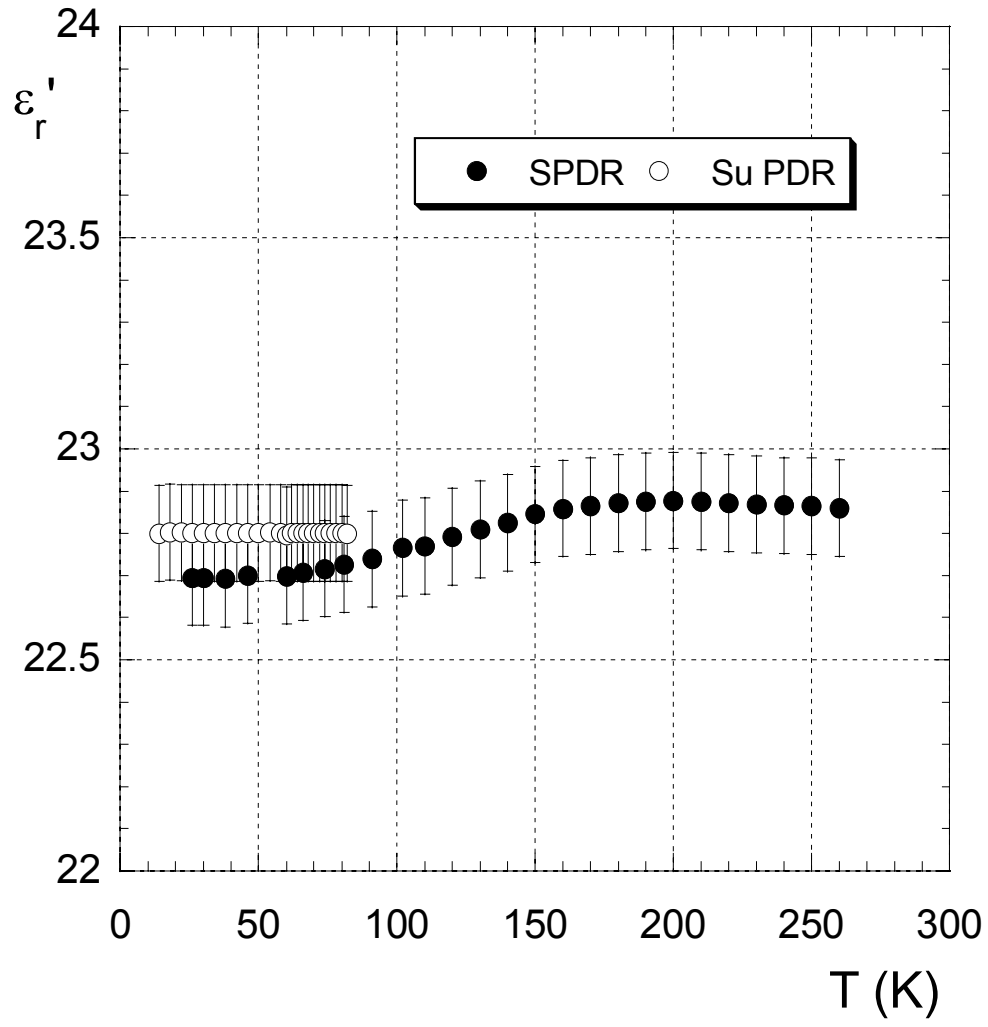

Fig. 5. Permittivity of LSAT substrate vs temperature

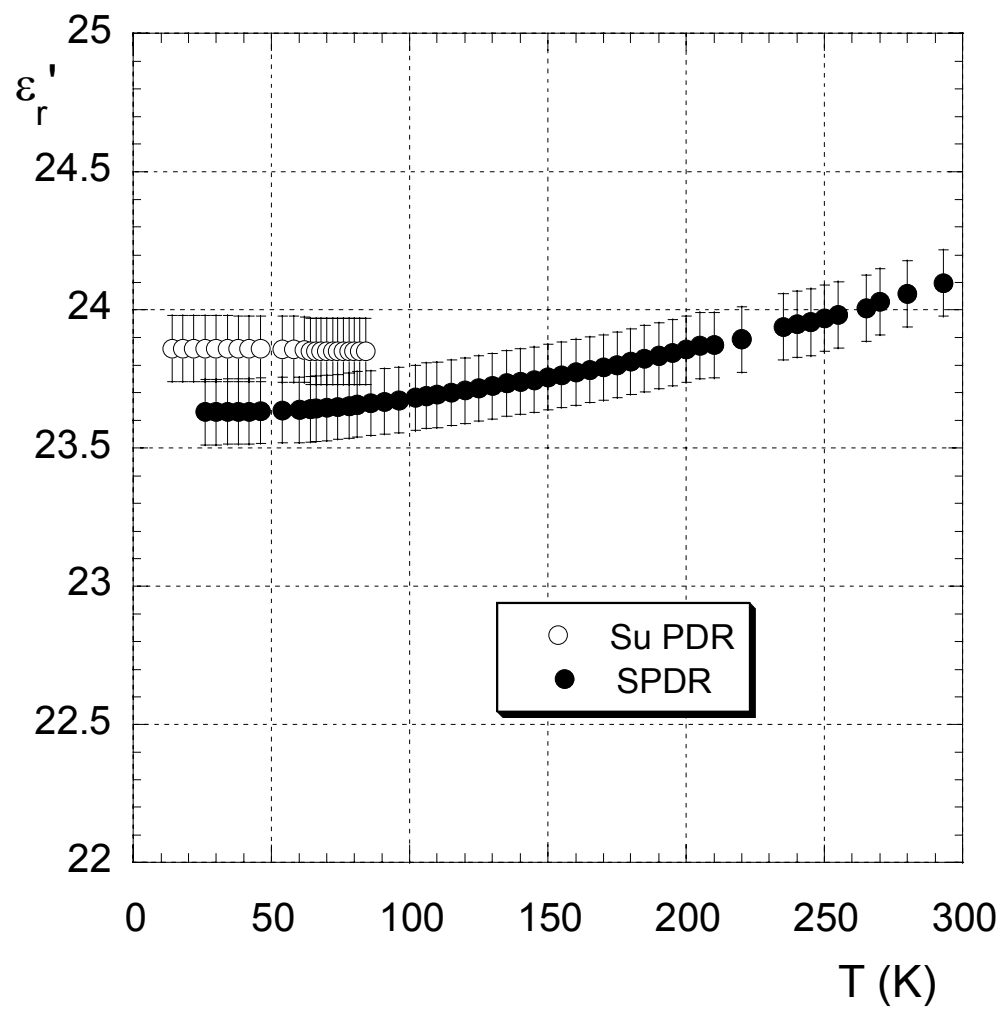

Fig. 6. Permittivity of $LaAlO_3$ vs temperature

For LAO and MgO the permittivity increases monotonically with temperature in contrary to LSAT whose permittivity reaches a plateau at temperature of 200K. Results of measurements of $\varepsilon_r$' presented in this paper agree well with results of bulk samples of the same dielectrics but in bulk form, employing the dielectric resonator technique [10, 11]. One can observe a discrepancy between values of $\varepsilon_r$' measured by the SPDR and the Su PDR techniques of approximately less than 0.5%. This can be caused by a systematic error associated with uncertainties in dimensions and dielectric properties of the dielectric rods and other parts of the resonators.

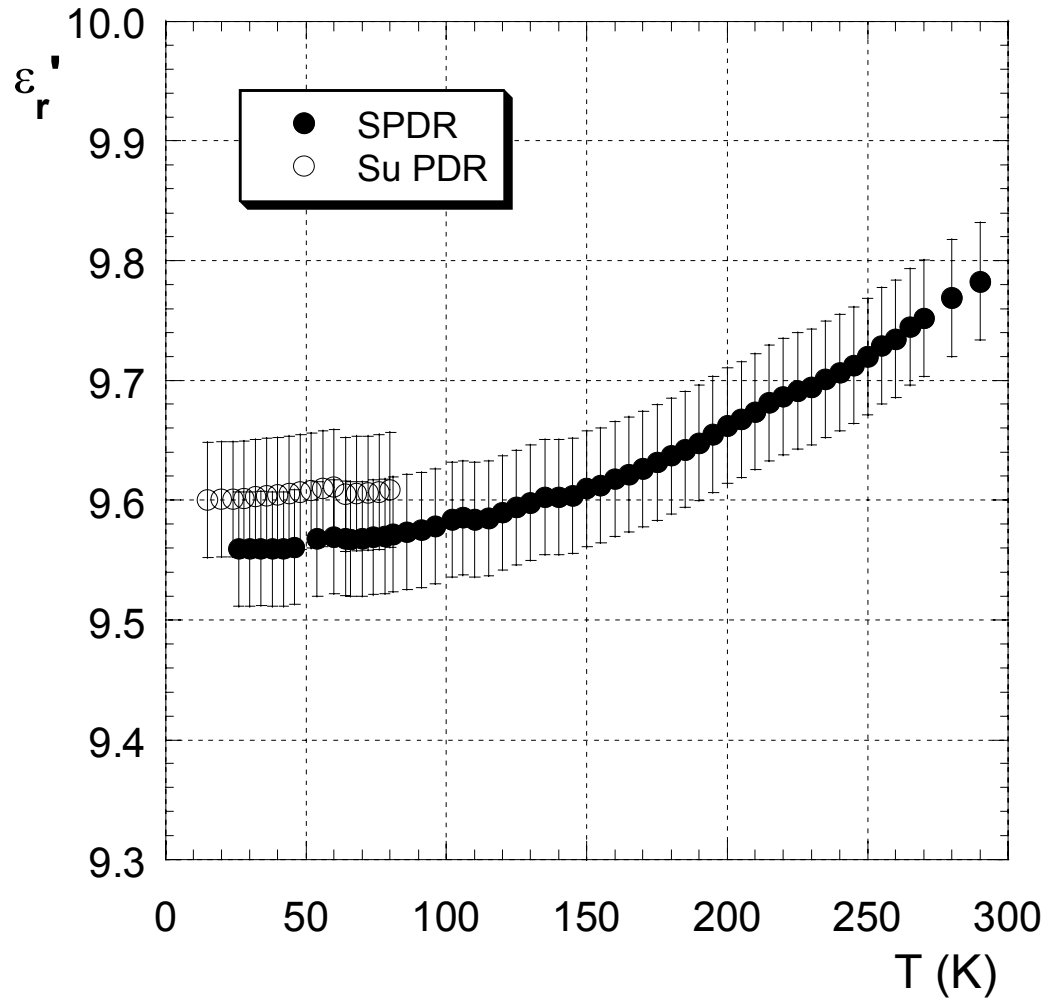

Fig.7. Real permittivity of MgO vs temperature

Measurement results of loss tangent obtained with both resonators for the LSAT substrate is shown in Fig. 8. A good agreement with discrepancies below 13% was obtained for temperatures higher than 50K. This is due to relatively large losses of the LSAT samples so it was possible to measure them accurately with both resonators**.**

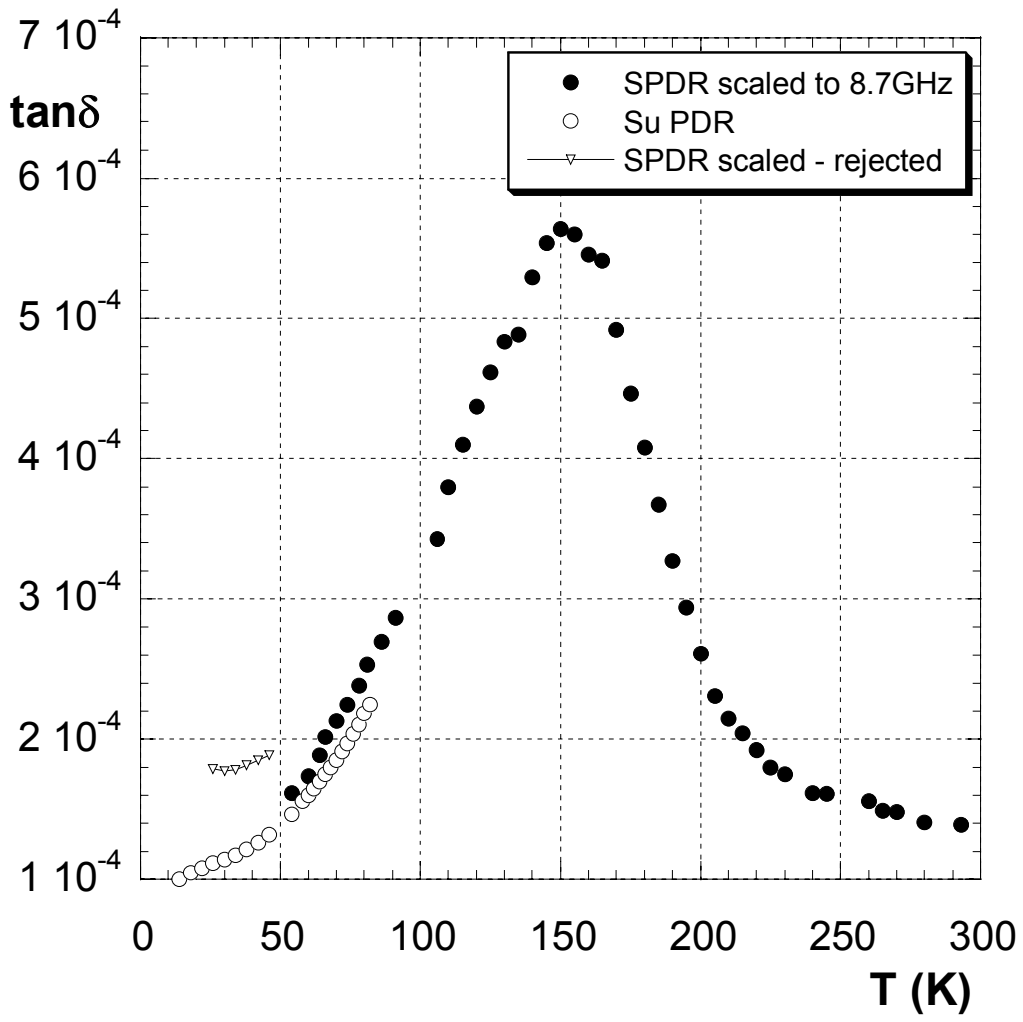

Fig.8. Loss tangent of LSAT substrate at 8.7 GHz

TABLE I

PERMITTIVITY AND DIELECTRIC LOSS TANGENT OF VARIOUS SUBSTRATES AT 77 K

| Material | $\varepsilon_r$ (77K) (SPDR) | $\varepsilon_r$ (77K) (Su PDR) | tanδ (SPDR) | tanδ (77K) (Su PDR) |
|---|---|---|---|---|
| LSAT | 22.72 ± 0.5% | 22.81 ± 0.5% | $2.4\times10^{-4}$ ± 20% | $2.05\times10^{-4}$ ± 5% |
| LaO | 23.65 ± 0.5% | 23.70 ± 0.5% | $4\times10^{-5}$ ± 50% | |
| MgO | 9.57 ± 0.5% | 9.61 ± 0.5% | $2\times10^{-5}$ ± 100% | $2\times10^{-6}$ ± 100% |
| Quartz | 4.41 ± 0.5% | 4.44 ± 0.5% | $3\times10^{-5}$ ± 100% | $2\times10^{-5}$ ± 20% |

Values of tanδ measured with SPDR below 50Kwere not confirmed by the Su PDR and hence these results were rejected. For lower loss materials including MgO and single crystal quartz losses are too small to be measured using the SPDR. For these two materials it was possible to assess the upper bound only for dielectric loss tangents. Summary of measurements of various materials at 77K is presented in Table I.

## III. CONCLUSIONS

The fabricated Split post resonator has proved to be useful for microwave characterization of low loss planar dielectrics in a wide temperature range from 20K to 400K. Measurements of real part of permittivity using this fixture can be performed with ±0.5% uncertainty and of loss tangent with the resolution of $2\times10^{-5}$ at the lowest temperature. The SPDR was used for measurements of the perpendicular components of $\varepsilon_r$' and tanδ of several low loss dielectric substrates in combination with the single post resonator with superconducting plates. The Su PDR provided increased resolution of dielectric loss tangent measurements of $2\times10^{-6}$ at temperatures below 80 K enabling verification (and elimination of some results) of the wide temperature measurement fixture. A good agreement with published results of bulk samples was obtained for $(La,Sr)(Al,Ta)O_3$, $LaAlO_3$, MgO and quartz substrates.

ACKNOWLEDGEMENT: This work was done under the financial support of ARC Large Grant A00799 (James Cook University). MVJ also acknowledges the JCU CRIG.